\documentclass[12pt,preprint]{aastex}

\usepackage{emulateapj5}
\usepackage{aastexug}
\usepackage{amsmath}
\usepackage{graphicx}
\usepackage{subfigure}
\usepackage{apjfonts}

\begin{document}
\def\gapprox{\mathrel{\vcenter{\offinterlineskip \hbox{$>$}
    \kern 0.3ex \hbox{$\sim$}}}}
\def\lapprox{\mathrel{\vcenter{\offinterlineskip \hbox{$<$}
    \kern 0.3ex \hbox{$\sim$}}}}

\newcommand{\Dt}[0]{\bigtriangleup t}
\newcommand{\Dx}[0]{\bigtriangleup x}
\newcommand{\E}{\mathcal{E}}
\newcommand{\D}{\bigtriangleup}
\newcommand{\beq}{\begin{equation}}
\newcommand{\eeq}{\end{equation}}
\newcommand{\mm}[2]{\textrm{minmod}\left({#1},{#2}\right)}
\newcommand{\sign}{\textrm{sign}}
\newcommand{\nf}{\mathcal{F}}

\title{Anisotropic winds from close-in extra-solar planets}

\author{James M. Stone}
\affil{Department of Astrophysical Sciences, Princeton University, Princeton,
NJ 08544}
\author{and Daniel Proga}
\affil{Department of Physics, University of Nevada, Las Vegas, NV 89154}

\begin{abstract}
We present two-dimensional hydrodynamic models of thermally driven winds
from highly irradiated, close-in extra-solar planets.  We adopt a very
simple treatment of the radiative heating processes at the base of the
wind, and instead focus on the differences between the properties of
outflows in multidimensions in comparison to spherically symmetric models
computed with the same methods.  For hot ($T \gapprox 2\times10^{4}$~K)
or highly ionized gas, we find strong (supersonic) polar flows are
formed above the planet surface which produce weak shocks and outflow
on the night-side.  In comparison to a spherically symmetric wind with
the same parameters, the sonic surface on the day-side is much closer
to the planet surface in multidimensions, and the total mass loss rate
is reduced by almost a factor of four.  We also compute the steady-state
structure of interacting planetary and stellar winds.  Both winds end in a
termination shock, with a parabolic contact discontinuity which is draped
over the planet separating the two shocked winds.  The planetary wind
termination shock and the sonic surface in the wind are well separated,
so that the mass loss rate from the planet is essentially unaffected.
However, the confinement of the planetary wind to the small volume
bounded by the contact discontinuity greatly enhances the column density
close to the planet, which might be important for the interpretation of
observations of absorption lines formed by gas surrounding transiting
planets.

\end{abstract}


\section{Introduction}

A significant fraction (about 20\%) of extra-solar giant planets (EGPs)
discovered to date have an orbit with a semi-major axis of less than 0.1
AU (Schneider 2008).  For such close-in EGPs, heating of the upper layers
of the atmosphere by irradiation from the central star, especially in
the extreme ultraviolet (EUV), can produce an extended envelope of gas,
and perhaps even drive a wind (e.g.  Moutou et al. 2000; see Ehrenreich
2008 for a review).  Direct detection of an extended envelope surrounding
HD209458b has been reported (Vidal-Madjar et al. 2003; 2008; Ehrenreich et
al. 2008), based on the absorption of stellar Ly-$\alpha$ during transits.
The fraction of the stellar Ly-$\alpha$ flux absorbed by HD209458b is
so large that it must be surrounded by a cloud of neutral hydrogen that
extends beyond the Roche lobe of the planet, and therefore is unbound
(although see Ben-Jaffel 2007 for an alternative analysis of the same
observations).  The observations also place a lower limit on the mass loss
rate from the planet of $\dot{M} \gapprox 10^{10}$~g~s$^{-1}$, although
the actual rate could be much higher since the absorption is saturated.

Detailed theoretical models of winds from highly irradiated EGPs are
of interest not only to interpret the observations, but also to place
firmer constraints on the mass loss rates.  If these rates are a factor
of few hundred times higher than the lower limit observed in HD209458b,
then over its lifetime the planet will lose a significant enough fraction of
its total mass to alter its structure and evolution (Hubbard et al. 2007;
Lecavelier des Etangs 2007).

A useful measure of the strength of the wind expected from a highly
irradiated EGP is the ratio of gravitational potential to thermal energy
at the top of the atmosphere, usually termed the hydrodynamic escape
parameter, that is
\begin{equation}
\lambda = \frac{GM_{p} \mu}{R_{p}kT}
\end{equation}
where $M_{p}$ and $R_{p}$ are the mass and radius of the planet, and $T$
and $\mu$ are the temperature and mean mass per particle in the atmosphere.
For $\lambda \gg 1$, the atmosphere is tightly bound and a hydrodynamic
wind is not expected, although a weak outflow may still be produced by
a variety of non-thermal processes, e.g. Hunten (1982).  For $\lambda
\lapprox 10$, a thermally-driven hydrodynamic (Parker) wind will be
produced (for reference, $\lambda \approx 15$ for $10^{6}$~K plasma
in the solar corona at 2$R_{\odot}$).  When evaluated at the effective
temperature $T_{\rm eff} \approx 10^{3}$~K of HD209458b, $\lambda \approx
140$, indicating a hydrodynamic wind is unlikely.  However, it has been
pointed out the upper layers of close-in EGPs will be heated to $T \sim
10^{4}$~K by the intense EUV radiation from the central star (Lammer et
al. 2003; Yelle 2004; see Ballester et al. 2007 for possible observational
detection of this hot gas).  At this temperature, and if the gas is mostly
neutral, $\lambda \approx 14$
(a value similar to that in the solar corona) so that
a thermally-driven wind is possible.  On the other hand, if the gas is
ionized (reducing $\mu$), or if the temperature is slightly higher
(both of which are relevant for EGPs around YSOs or higher mass stars), then
$\lambda$ can be even smaller.

Calculating the structure and mass loss rate from a thermally-driven
hydrodynamic wind from close-in EGPs requires solving the time-dependent
equations of hydrodynamics, with a proper accounting of the radiative
heating at the base of the wind.  The latter is probably the most
challenging aspect of the problem, and different approaches have led to
different estimates of the mass loss rates.  To simplify the calculation,
all of the models presented to date are one-dimensional (spherically
symmetric).  Watson et al. (1981) presented models for the escape of
hydrogen from the terrestrial planets in the early solar system, using
a simplified treatment of the heating that deposits all of energy in a
single zone at the base of the wind.  Scaled-up versions of these models
have been applied to EGPs.  Tian et al (2005) improved these solutions
by using a multi-dimensional radiative transfer calculation to estimate
the radial distribution of the heating rate (the underlying hydrodynamic
models are still 1D, however).  Lecavelier des Etangs et al. (2004)
pointed out that tidal forces could distort the gravitational potential
isosurfaces around the planet, and this could affect the mass loss rate
(Erkaev et al. 2007), but only if the shape and location of the sonic
surface are significantly altered.  Yelle (2004) has presented more
realistic models of the wind from close-in EGPs including chemistry,
photo-ionization and recombination, and thermal and molecular diffusion
in a hydrodynamic model (again assuming spherical symmetry).  New models
with a similar level of sophistication have recently been computed for
HD209458b by Garc\'{i}a Mu\~{n}oz (2007), and for EGPs in general by
Murray-Clay et al. (2008).

As has been pointed out by many authors, the fact that close-in EGPs
are irradiated on only one side calls into question the assumption
of spherical symmetry.  The problem is exasberated by the fact that
close-in EGPs will be tidally locked, so that the only mechanisms that
can transport heat to the night-side are bulk flows in the atmosphere
and thermal conduction.
Although global circulation models are beginning to be investigated for
close-in EGPs (e.g. Dobbs-Dixon \& Lin 2008; Showman et al. 2008),
these models do not apply to
the weakly bound hot ($T \sim 10^{4}$~K) gas in the upper atmosphere.
For HD209458b, the advection time for gas moving at the sound speed to
move half way around the planet is $\approx 11$ hours, much longer than
the radiative cooling time for the hot gas.  Thus, there is likely to
be a large contrast in the temperature between the day and night sides
in the upper layers of the atmosphere which serve as the base of the wind,
even if the temperature at the infrared photosphere is more uniform.
The amplitude of this contrast is, however, uncertain.

The goal of this paper is to investigate the effect that anisotropic
heating has on the winds from close-in EGPs using two-dimensional
numerical hydrodynamic calculations.  Given the uncertainties in the
mass loss rate produced by different treatments of the radiative heating
and microphysics in the wind, we adopt the simplest possible approach,
and focus on the {\rm relative} difference between our multidimensional
calculations and spherically symmetric models computed with identical
techniques.  We find there are important differences in multidimensions.
For example, the sonic surface in the wind is moved much closer to the
planet, and the overall mass loss rate is reduced by almost a factor of
four for a nearly isothermal wind.  Our numerical approach also allows
us to consider the interaction between planetary and stellar winds.
Although a stellar wind does not affect the mass loss rate, it confines
the planetary wind to a small volume and greatly enhances the column
density close to the planet.  These results suggest that incorporating
multidimensional effects may be as important as improved treatments
of the radiative transfer and microphysics in order to interpret the
observations of transiting planets such as HD209458b.

In addition to a simplified treatment of the thermodynamics, we have made
a number of other simplifying assumptions in this first investigation.
For example, our calculations are two-dimensional (axisymmetric), which
precludes us from studying the effect of Coriolis forces (orbital motion)
on the outflow.  Therefore, our models are appropriate only for the flow
close to the planet.  Fully three-dimensional hydrodynamic models of an
{\em isotropic} wind from orbiting planet have recently been presented
by Schneiter et al. (2007).  Such calculations are challenging, even
with an adaptive mesh these authors could only afford 3 grid points per
planetary radius in the model.  In additions, our models are hydrodynamic
rather than MHD, even though at large radii the wind may be significantly
ionized, and interact with a primarily MHD stellar wind.  Finally, there
are a number of kinetic plasma effects (e.g. charge-exchange reactions
between the planetary and stellar wind particles) that might become
important as the wind becomes very diffuse at large radii.  These effects
are known to be important in the MHD of the heliopause (e.g. Borovikov
et al. 2008).  Fully 3D MHD models of anisotropic winds from planets
including multidimensional radiative transfer, microphysics, and the
appropriate kinetic plasma effects are an interesting and important
direction for future work.

The organization of this paper is as follows.  In the next section,
we describe our numerical methods.  In \S3, we present results for
anisotropic winds from isolated EGPs.  In \S4, we consider the interaction
of anisotropic winds from an EGP with a stellar wind.  Finally, in \S5
we discuss our results, and in \S6 conclude.

\section{Method}

We use the time-dependent hydrodynamics code ZEUS (Stone \& Norman 1992)
to compute the multidimensional structure of an EGP wind.  We start
our calculations from an initially spherically symmetric outflow,
and run them for many crossing times, until the solution has settled
into a steady-state.  Typically this takes less than an hour on a
modern workstation.  This technique is much simpler than solving the
steady-state equations directly, even in 1D, because the steady-state
ODEs define a two-point boundary value problem which contains a critical
point that requires special treatment when solved using shooting or
relaxation methods.

We solve the hydrodynamic equations in spherical polar $(r,\theta)$
coordinates, that is
\begin{equation} 
\frac{\partial \rho}{\partial t} + 
\frac{1}{r^{2}}\frac{\partial (r^{2}\rho v_{r})}{\partial r} +
\frac{1}{r\sin\theta}\frac{\partial (\rho v_{\theta} \sin\theta)}{\partial \theta} = 0
\label{eq:cons_mass}
\end{equation}
\begin{equation} 
\rho \left( \frac{\partial v_{r}}{\partial t} +
v_{r}\frac{\partial v_{r}}{\partial r} +
\frac{v_{\theta}}{r}\frac{\partial v_{r}}{\partial \theta} -
\frac{v_{\theta}^{2}}{r} \right) = - \frac{\partial P}{\partial r} -
\frac{GM_{p}}{r^{2}}
\label{eq:radial_velocity}
\end{equation}
\begin{equation} 
\rho \left( \frac{\partial v_{\theta}}{\partial t} + 
v_{r}\frac{\partial v_{\theta}}{\partial r} + 
\frac{v_{\theta}}{r}\frac{\partial v_{\theta}}{\partial \theta} +
\frac{v_{r} v_{\theta}}{r} \right) = -\frac{1}{r}\frac{\partial P}{\partial \theta}
\label{eq:angular_velocity}
\end{equation}
\begin{equation} 
\frac{\partial e}{\partial t} +
v_{r}\frac{\partial e}{\partial r} +
\frac{v_{\theta}}{r}\frac{\partial e}{\partial\theta}  = 
- \frac{P}{\rho} \left(\frac{1}{r^{2}}\frac{\partial (r^{2}v_{r})}{\partial r} +
\frac{1}{r\sin\theta}\frac{\partial(v_{\theta} \sin\theta)}{\partial \theta} \right) 
\label{eq:energy}
\end{equation}
where $\rho$ is the mass density, $v_{r}$ and $v_{\theta}$ are the radial
and angular components of the velocity, and $e$ is the internal energy
density.  The pressure $P$ is related to $e$ through the equation of
state, $P=(\gamma -1)e$, which implies $T=(\gamma -1)\mu e/k\rho$.
We also add an artificial viscous stress term to the momentum and
energy equations to ensure proper shock capturing, see Stone \& Norman
(1992) for the details of these terms in spherical polar coordinates.
The axis of symmetry of the grid is orientated along the line connecting
the centers of the planet and the central star, with $\theta=0$ the
direction towards the star (noon), and $\theta=\pi$ the direction away
from the star (midnight).

Although we are solving the dynamical equations in a frame co-rotating
with the planet, we have neglected the Coriolis force terms in equations
\ref{eq:radial_velocity} and \ref{eq:angular_velocity}.  Since we find
that the terminal velocity $V_{\infty}$ of the wind is less than the
orbital velocity $V_{\rm orb}$ of an EGP at 0.1 AU from a solar type star
(typically, we find $v_{\infty} \sim V_{\rm esc} < V_{\rm orb}$ where $V_{\rm
esc} \approx 40$~km/s is the escape velocity from the planet), then the
neglect of Coriolis forces limits the applicability of our solutions to
regions close to the EGP.  On larger scales (on order of the orbital radius),
the planetary wind will be deflected by Coriolis forces, and swept back
by the stellar wind into a cometary shape (Moutou et al. 2001;
Vidal-Madjar et al. 2003).  We discuss physical effects that are likely to be
important for the structure of the wind on large scales in \S5.

Also note that we do not explicitly include heating, cooling, or
thermal conduction terms in the energy equation \ref{eq:energy}.
Instead of injecting a fixed rate of heating at the base of the wind,
and then modeling the cooling and conduction processes that determine the
temperature, we simply fix the temperature directly.  Operationally, in the
first grid cell above the lower boundary condition at the surface of
the planet $r=R_{p}$, we hold the density at a constant $\rho_{0}$, and
set the internal energy to
\begin{equation}
 e(R_{p},\theta) =  e_{0} \max (0.01, \cos{\theta})
\label{eq:Tprof}
\end{equation}
where $e_{0} = \rho_{0}/[\gamma (\gamma-1)]\lambda_{0}$.  This introduces
the hydrodynamic escape parameter at the base of the wind $\lambda_{0}$
as a free parameter.  By holding
$\rho=\rho_{0}$ and $e=e_{0}$ in the first radial cell, the wind solution
emerges naturally from the density and pressure gradients above this cell.
Provided that the acceleration region between this cell and the sonic
point is well resolved, the flow (including the mass loss rate) is then
set self-consistently by the resulting density and pressure gradients.
This technique is identical to procedure we have used before to model
radiative driven winds in a disk geometry (e.g.  Proga, Stone, \& Drew
1998; Proga, Stone, \& Kalman 2000).  Typically we have at least
20 cells in the wind acceleration region below the sonic point.
The angular distribution we have assumed for the internal energy
(equation \ref{eq:Tprof}) gives a temperature ratio of 100 between the
day and night sides.  Since the amplitude of this ratio is uncertain,
we will also present results in the following section for a model in
which this ratio is only two.

We have also computed models in which the both the internal energy
and density are varied according to equation \ref{eq:Tprof}, so that
the temperature at the base of the wind is fixed (this may be a better
model of a photoionized atmosphere, where higher UV fluxes on the day
side produce ionization at higher densities, but produce little change
in the temperature).  Since the pressure distribution in the wind using
this method is identical to simply varying $e$, we find essentially no
change between models launched with fixed temperature or
fixed density at the base.

Equations \ref{eq:cons_mass} through \ref{eq:energy} are discretized
on a grid of 200 radial and 200 angular cells, in the domain $1 \le
r/R_{p} \le 50$ and $0 \le \theta \le \pi$.  We use a non-uniform grid
in the radial direction, with the size of each successive radial cell
increased by the ratio $\delta r_{i+1}/\delta r_{i} = 1.02$.  This gives
better resolution in the inner regions of the grid (at $r=R_{p}$,
$\delta r/R_{p} \approx 0.02$), and also keeps the cells nearly square
throughout the domain ($r\delta \theta \approx \delta r$).  The boundary
conditions at $\theta=0$ and $\pi$ are given by symmetry conditions.
At the outer radial edge of the grid, we use outflow boundary conditions
(all variables projected at zero slope).  At the inner radial edge,
we use a reflecting boundary condition.

We initialize the grid above the first radial cell to contain extremely
low density ($\rho = 10^{-20}\rho_{0}$) gas within an internal energy
profile $e(r,\theta) = e_{0}/r^{2}$, a uniform outward radial velocity
$v_{r}=C_{s}$, where $C_{s}^{2}=\gamma(\gamma-1)e_{0}/\rho_{0}=1/\lambda_{0}$,
and no angular motion $v_{\theta}=0$.  The outward flow helps to prevent
gravitational infall developing in the outer regions, and the low density
helps to reduce transients.  The flow crossing time across the entire grid
is $t \sim L/C_{s} = 50$, where $L$ is the radius of the outer boundary.
We find by $t \approx 100$, a steady solution has been established.
All of the results presented in this paper are at a time of $t=200$.
We have confirmed that the steady state wind solution is independent
of how we initialize the wind, and is converged with the numerical
resolution used here (quantities such as the mass loss rate and terminal
velocity change by less than 1\% if the numerical resolution is halved
in each direction).

The primary parameters that determines the wind solution are the
adiabatic index $\gamma$ and $\lambda_{0}$ at the base of the wind.
We discuss wind solutions for a variety of values for $\gamma$, from
$\gamma=1.01$ (nearly isothermal) to $\gamma=5/3$ (adiabatic).  We use
$\lambda_{0}=5$ for most of the models presented in this study, which
implies $\lambda=500$ on the night side.  The value $\lambda_{0}=5$
is three times smaller than the value for HD209458b, assuming the gas
temperature is $T=10^{4}$~K, and that it is mostly neutral.  Choosing a
smaller value of $\lambda_{0}$ is mostly a numerical convenience: higher
values lead to much lower density winds in which transients are much more
severe, and therefore take much longer to reach a steady state.  In fact,
the value $\lambda_{0}=5$ is relevant to EGP winds which are ionized (and
therefore have smaller $\lambda$ for the same temperature), or in which
the temperature is slightly (a factor of two) larger.  Both cases are
relevant to EGPs around YSOs with higher UV fluxes, or which orbit higher
mass central stars.  In order to investigate the structure of winds with
higher $\lambda_{0}$, we will also present a model with $\lambda_{0}=10$.

To test our numerical methods, and to compare to previous models, we
have calculated a number of spherically symmetric models with different
values of $\lambda_{0}$ and $\gamma$.  Our method reproduces location of
the sonic point in 1D Parker winds (Keppens \& Goedbloed 1999) to better
than 1\%.  We have also recover the same structure and mass loss rates
for isothermal winds found by Watson et al. (1981) and Tian et al. (2005)
for both $\lambda=5$ and 15.

\section{Results}

We start by discussing the two-dimensional structure of a nearly
isothermal wind, $\gamma=1.01$, with $\lambda_{0}=5$.  Figure 1 shows
images of the density (with velocity vectors overlaid) and temperature
in the inner regions $r/R_{p} \le 10$ after the flow has relaxed to a
steady-state.  Remarkably, even though $\lambda=500$ on the night-side
of the planet, far too large to drive an appreciable wind, the density
in the outflow near $\theta=\pi$ is quite large.  This indicates there
must be a substantial non-radial flow around the planet to maintain
a significant density on the night-side.  Contours of the density are
slightly elongated in the polar directions compared to the region near
$\theta=\pi/2$, by about a factor of 5/4, indicating the density in the
wind is mildly anisotropic.

A prominent feature in the density is a discontinuity that starts at
$\theta \approx 3\pi/4$ near the planet, and curves to smaller angles as
it extends outward in radius.  This feature is a shock front produced by
geometrical compression of the wind as it flows from the day-side to the
night-side of the planet, and is further indication of a strong non-radial
motions in the outflow.  The polar flow can be seen more clearly in the
pattern of velocity vectors.  Close to the planet (at $r \le 2R_{p}$), the
velocity in the wind near the region near $\theta=\pi/2$ is nearly purely
polar, while on the night-side there is {\em inflow} rather than outflow.
At large radii, there is a nearly perfectly radial outflow at all angles,
although the length of the velocity vectors clearly shows the terminal
velocity is larger on the day-side in comparison to the night-side.
We investigate the velocity field of the wind more thoroughly below.

Since for $\gamma=1.01$ the wind is nearly isothermal, the temperature
is not significantly affected by geometrical compression or expansion.
The temperature in the wind essentially becomes a passive scalar which
is frozen-in to the value set at the surface of the planet, and contours
of the temperature are very nearly equal to the streamlines in the flow.
Thus, the temperature image in Figure 1 demonstrates how the outflow
curves around the planet's surface.  The temperature in the night-side
flow is everywhere constant and equal to the temperature at the base
of the wind emanating from $\theta\approx \pi/4$ on the day-side, about
$\log T \approx -0.85$ in the units used in the figure. There is a small
region of lower temperature very close to the planet near the terminator,
but at $\theta > 3\pi/4$, the temperature is once again equal to the
value elsewhere on the night side, $\log T \approx -0.85$.  This is an
indication of infall on the night side.

Overlaid on the temperature image are contours of the sonic surface,
i.e. regions where $\vert v \vert /C_{s}=1$, where $v$ is the total
magnitude of the velocity, including both the radial and polar components.
On the day-side, the sonic surface is located at about $0.5R_{p}$
above the surface of the planet, and this decreases to zero at the
terminator, as expected.  The contours on the night-side of the planet
indicate a region where the wind is {\em decelerated} to subsonic
flow by the shock front identified in the density image.  The
flow is then re-accelerated to supersonic outflow at $r \gapprox 5R_{p}$
(for $\theta=\pi$).  The contour very near the surface of the planet
on the night-side indicates supersonic infall there.  For comparison,
the sonic surface for a spherically-symmetric wind with $\lambda_{0}=5$
is located at roughly $1.5R_{p}$ above the surface of the planet.  Thus,
the sonic surface is moved substantially closer to surface of the planet
in the anisotropic case.

Figure 2 shows radial slices of the density and radial Mach number,
$v_{r}/C_{s}$, at $\theta=0$, $\pi/2$, and $\pi$.  Also shown for
comparison is the spherically symmetric wind for the same value of
$\lambda_{0}$.  The difference in the density in the outflow between
the day-side and the night-side is large only very close to the planet;
near $r=2R_{p}$ the density on the day-side can be nearly an order
of magnitude larger than on the night-side.  This difference drops
to less than a factor of two at $10R_{p}$, and becomes even smaller
farther out.  However, the density at all angles is much less than the
spherically symmetric case, by a factor of close to 4.  The radial Mach
number shows the flow speed is faster than the spherical case on the
day-side, about the same at the terminator, and significantly slower on
the night-side.  Infall on the night-side is clearly evident by the
negative values for the Mach number below $r/R_{p} \approx 3$.  An important
quantity is the mass loss rate in the wind.  For this model we find the
total (angle-integrated) steady-state mass loss rate is a factor of 3.7
smaller than the spherically symmetric case.

The non-radial flow in the wind can be explored by one-dimensional
slices in polar angle at different radii.  Figure 3 shows slices of the
density and both components of the velocity (scaled to the sound speed)
taken at $r/R_{p}=2$, 5, 10, and 50.  The amplitude of the spherically
symmetric solution at these locations is also shown for each quantity.
The shock front evident in the density image (figure 1) is most prominent
in the polar velocity (bottom panel of figure 3).  At small radii,
there is a large increase in the polar velocity moving from the day
to night sides, until there is a sudden, discontinuous drop at $\theta
\approx 3\pi/4$.  This trend is repeated in each slice at larger radii,
except the maximum amplitude of the polar velocity drops, and the shock
front is moved to smaller angles.  At $r/R_{p}=2$, the polar velocity is
supersonic, reaching a maximum Mach number of about 1.5.  The location of
the shock is also clearly evident in the angular profiles of the density.
At each radius there is a slow decrease in the density moving from the
day-side to the night-side, with a discontinuous jump at the location
of the shock front.  The shock is weak, it produces a density jump of
only about 2.5 at $r/R_{p}=2$.  The radial velocity is not strongly
affected by the shock, a reflection of the fact that this component
of the velocity is almost parallel to the nearly radial shock front.
Instead, the primary feature of the radial velocity is a systematic
decrease from the day-side to the night-side.

To explore the effect of varying $\gamma$ on the two-dimensional structure
of the wind, figure 4 shows images of the density and temperature (with
velocity vectors and the sonic surface overlaid) from a model computed
using $\gamma=1.1$, but otherwise identical to the nearly isothermal model
discussed above.  Comparison of figures 1 and 4 show that the density
structure of the wind is very similar in the two cases.  The polar flow
towards the night side of the planet is very clear in the pattern of
the velocity vectors in figure 4.  Once again, the length of the vectors
demonstrate that the outflow velocity is much higher on the day-side in
comparison to the night-side.  The temperature profile in the wind for
$\gamma=1.1$, however, is quite different than that for $\gamma=1.01$.
Now, adiabatic expansion produces significant cooling in the wind, so the
temperature at large radii drops dramatically.  Moreover, the shock which
decelerates the polar flow towards the night side produces a significant
jump in temperature, and so is clearly visible in the right panel of
figure 4.  Due to the slower acceleration of the wind, the sonic surface
is moved to larger radii, it is now at about $0.8R_{p}$ above the surface
of the planet.  This is still much smaller than the spherically symmetric
case, where for $\gamma=1.1$ the sonic surface is about $3R_{p}$ above
the planet surface.  The mass loss rate in this case is about 2.9 times
smaller than the mass rate in a spherically symmetric wind with the same
$\lambda_{0}$ and $\gamma$, and this value is 2 times smaller than the
mass loss rate in the anisotropic wind with $\gamma=1.01$ discussed above.

As $\gamma$ is increased further, the general structures noted above
remain the same, however the outflow becomes systematically weaker.
For $\gamma$ close to 5/3, a thermally driven wind is not possible,
and we find no steady-state outflow (Keppens \& Goedbloed 1999).

We have also computed a model in which the minimum internal energy on the
night side in equation \ref{eq:Tprof} is $0.5e_{0}$, giving a temperature
ratio of two between the day and night sides.  We use $\gamma=1.01$ and
$\lambda_{0}=5$
to allow direct comparison to the model discussed above.  As expected,
the flow in this case is more nearly spherically symmetric.  However,
important features such as the location of the sonic surface are little
changed from the result shown in figure 1, so that the total mass loss
rate is nearly identical to the model with a larger temperature contrast.
Since the details of the wind depend on the temperature contrast between
the day and night sides, determining this ratio at the base of the wind
(as opposed to the infrared photosphere) from observations is important.

Finally, we have also explored the effect of increasing $\lambda_{0}$
on the properties of the wind.  Figure 5 presents radial slices of the
density and radial Mach number, $v_{r}/C_{s}$, at $\theta=0$, $\pi/2$,
and $\pi$.  Also shown for comparison is the spherically symmetric
wind for the same value of $\lambda_{0}$.  The density at all angles is
much smaller (more than an order of magnitude) than the $\lambda_{0}=5$
case presented in figure 2.  As before, the difference in the density
in the outflow between the day-side and the night-side is not large.
Moreover, the density at all angles is not significantly different than
the spherically symmetric case, it is only a factor of 0.8 smaller.  Thus,
for this large value of $\lambda_{0}$, anisotropic effects are reduced.
The plot of the radial Mach number indicates the reason why: the sonic
surface of the wind is moved to much larger radii, about $3R_{p}$.
Two-dimensional images of the flow show that the density below the sonic
surface is nearly spherically symmetric.  There is a very strong, nearly
supersonic polar flow below the sonic surface
that transports heat and mass to the night side.  The region
below the sonic surface forms
an extended atmosphere which serves as the base of a nearly spherically
symmetric wind.  While the difference in the flow pattern in this case
is of interest, it may not be realistic due to the simplified treatment
of the thermodynamics.  The polar circulation (which is analgous to the
flow in a GCM) that produces a spherical wind is likely strongly affected
by the local heating and cooling processes, which we have not modeled
directly here.  Thus, we conclude that the structure of EGP winds in
multidimensions can be affected by the value of $\lambda_{0}$ at the
base of the wind, and that models with large values of $\lambda_{0}$
require more realistic treatments of radiative heating and cooling.
Since larger values of $\lambda_{0}$ are likely relevant to many
observed EGPs, exploring this regime with more realistic
multi-dimensional models is important.

\section{The Interaction of Planet and Stellar winds.}

Using the numerical methods adopted for this study, it is straightforward
to compute the interaction of an anisotropic wind from a close-in
EGP with the wind from the central star.  As an initial condition,
we use the steady state structure of the fiducial wind model (with
$\gamma=1.01$) studied in the previous section, computed using $\lambda_{0}=5$.
We restart this model
with the outer boundary condition modified to represent an inflowing
stellar wind.  Specifically, for $\theta \le \pi/2$, we set the density
and internal energy at the outer boundary to be constants, $\rho_{*}$
and $e_{*}$ respectively, and set the radial and angular velocities
to be $v_{r} = v_{*}\cos{\theta}$ and $v_{\theta}=v_{*}\sin{\theta}$,
where $\rho_{*}=10^{-4}\rho_{0}$, $e_{*}=2.5 \times 10^{-3}e_{0}$, and
$v_{*}/C_{s,*}=-\sqrt{5}$, where $C_{s,*}$ is the sound speed in the
stellar wind.  We then follow the propagation of the stellar wind across
the grid, and stop the calculation when the structure resulting from its
interaction with the anisotropic EGP wind has settled into a steady state.

Note that for the parameter values adopted for the stellar wind,
$C_{s,*}^{2}/C_{s,0}^{2}=25$, that is the stellar wind is much hotter than
the planet wind.  For the properties of HD209458b, the stellar wind would
have a velocity of $\approx 90$~km/s and a temperature of $2.5\times
10^{5}$~K, both within the range of stellar wind properties expected
for EGPs (Preusse et al. 2005).  Note also that pressure driven stellar
winds do not reach their terminal velocity until many stellar radii, so
that a planet with an orbital radius of $\lapprox 0.1$~AU may still be
inside the wind acceleration region.  Furthermore, the stellar wind is
likely to be magnetized, so that the interaction between the two should
be MHD.  Finally, since the stellar wind velocity is comparable to the
orbital velocity of the planet, in the frame of reference of the planet
(in which our computations are performed), the stellar wind would arrive
from a direction offset by a large angle from the line connecting the
centers of the star and planet.  Thus, the interaction should ideally be
computed in full 3D MHD, however this is beyond the scope of this study.
Three-dimensional hydrodynamic models of the interaction of planet and
stellar winds have been presented by Schneiter et al. (2007).

Figure 6 shows images of the density (with velocity vectors overlaid)
and temperature after the interacting planetary and stellar winds have
settled into a steady-state.  The density image shows the planetary
wind is swept back into a parabolic-shaped region that opens towards
the night-side of the planet.  This region is bounded by a contact
discontinuity that separates the shocked planet and stellar winds.
A shock in the stellar wind upstream of this contact is clearly visible
as a discontinuity in the density.  The pattern of velocity vectors
show how this shock diverts the stellar wind around the region occupied
by the outflow from the planet.  A shock in the planet wind upstream
of the contact discontinuity is also clearly visible in the density
image at $r/R_{p}$ of between 3 and 3.5 for $0 \le \theta \le \pi/2$.
At large radii the density in the planetary wind is remarkably constant
with radius.   This is because the cross-sectional area of the parabolic
region occupied by the planet wind is nearly constant at large radii,
so that there is no further geometrical expansion in the planetary wind.
Thus, once the planetary wind reaches terminal velocity, its density
becomes constant.

The temperature image again mostly shows streamlines in the flow,
since the model is nearly isothermal.  In contrast to the flow shown in
figure 1, the flow in figure 6 clearly follows streamlines that strongly
curve away from the star in the wind interaction region.  Contours of
the sonic surface are overlaid on the temperature image in figure 6.
Moving outwards from the planet on the day-side, there are three contours.
The first is the location of the sonic surface in the planetary wind,
it is located about $0.5R_{p}$ above the planet surface.  The location
and shape of this surface is essentially identical to the case with no
stellar wind (figure 1).  The next contour, located at about $2.5R_{p}$
at $\theta=0$, is the planetary wind termination shock.  This contour
is identical to the shape of the shock visible in the density image.
Clearly, the termination shock {\em decelerates} the planetary wind to
subsonic velocities.  Finally, the last contour at about $4.75R_{p}$
above the planet surface at $\theta=0$, is the stellar wind termination
shock.  Note this contour only follows the location of the stellar
wind shock visible in the density for angles $\theta \lapprox \pi/4$.
This indicates the transverse velocity becomes supersonic as the stellar
wind is diverted around the planet.

It is important to note that the planetary wind sonic surface and the
planetary wind termination shock are well separated, by about $2R_{p}$,
over most of the flow.  Thus the sonic surface in the planetary wind
is little changed by the interaction with the stellar wind; and the
stellar wind has little effect on the mass loss rate from the planet.
Of course, the exact location of the wind termination shocks and the
contact discontinuity separating the two depends on the assumed momentum
flux in the stellar wind.  More powerful winds would confine the planetary
wind to a region closer to the planet.  However, since the momentum flux
density in the planet wind increases rapidly near the planet, it would
be difficult for the stellar wind to reach the sonic surface of the
planetary wind, unless the hydrodynamic escape parameter $\lambda_{0}$
were considerably larger.

The radial profiles of the density at angles of $\theta=\pi/2$ (terminator)
and $\pi$ (away from star) are shown in figure 7.  For comparison, the
profile for a spherically symmetric wind with $\lambda_{0}=5$ is plotted
as a solid line.  The profile of the density at the terminator (shown as a
dashed line) is dominated by a series of steps representing the shocks
in the planet and stellar winds, and the contact discontinuity that
separates them.  Moving outward from the planet, the density decreases
sharply in the wind acceleration region, then increases discontinuously
at the wind termination shock at $r/R_{p} \approx 2$.  It then drops
discontinuously at the contact discontinuity between the shocked winds
at $r/R_{p} \approx 6$, and drops again discontinuously at the stellar
wind shock at $r/R_{p} \approx 10$.  Beyond this radius the density
is constant, as this region is filled with unshocked stellar wind.
In contrast, the slice along the night side (shown as a dot-dashed
line) does not cross the boundary between the planet and stellar winds.
Thus, in this case the density drops smoothly from the planet surface
outwards, reaching a constant value by $r/R_{p} \approx 10$.  This is in
stark contrast to the density profile in the wind from an isolated planet
(see figure 2), which declines at all radii.  At $r/R_{p} = 10$, the
density in this case is over ten times larger compared to an anisotropic
wind with no stellar wind interaction; at $r/R_{p} = 50$ it is nearly
1000 times larger.  Despite this difference, the mass loss rate in the
planetary wind is nearly identical to the case with no stellar wind,
because the conditions in the planetary wind at the sonic surface are
relatively unaffected.  However, the increased density in the planet
wind strongly affects the observed column density through the wind,
as we discuss in the next section.

\section{Discussion}

Perhaps the most direct observable of winds from close-in EGPs is the
column density of gas surrounding the planet observed during transits.
It is of interest to investigate whether the column density of anisotropic
winds, and anisotropic winds that are confined by the interaction with a
stellar wind, are significantly different from spherically symmetric
models.

Figure 8 shows profiles of the column density (in arbitrary units) as a
function of impact parameter for a spherical EGP wind, an anisotropic
EGP wind, and an anisotropic wind that interacts with a stellar wind
as viewed from two different angles.  With no stellar wind, comparison
of the spherical and anisotropic winds shows that the lower mass loss
rate and density at large radii in the latter significantly reduces
the column density at a few planetary radii.  However, the situation is
reversed as soon as the interaction with a stellar wind is considered.
If the observer is located at $\theta=\pi$ (that is the viewing angle is
directly towards the night-side), then the column density is strongly
enhanced out to the radius of the contact discontinuity between the
planetary and stellar winds, because all of the planet wind material
is confined to this region.  If the interacting winds are viewed at an
angle of $20^{\circ}$ from $\theta=\pi$, then the region of large column
density associated with the compressed planetary wind is extended to one
side of the planet, and would be visible as a cometary tail extending
away from the planet.

These results suggest the interaction with a stellar wind can greatly
enhance the column density of close-in EGP winds.  However, there are some
limitations to our results.  For $r/R_{p} \gapprox 10$, the structure
of the planetary wind will be strongly affected by orbital motion and
tidal forces from the central star.  This will strongly affect the
column density at these radii; which warrants further 3D hydrodynamic
(Schneiter et al.  2007) and MHD investigations.  Moreover, we find the
terminal velocity of an anisotropic planetary wind is only 3-4 $C_{s}$,
which for $10^{4}$~K gas is only about $V_{\infty} \approx 30$~km/s.
This is much smaller than the escape velocity from a solar type star at
0.1~AU.  Thus, although the wind is unbound from the planet, it will not
escape the stellar gravitational potential, and will most likely collect
in a torus of gas at the orbital distance of the planet.  Whether the
stellar wind, or radiation pressure, or non-thermal effects such as
charge exchange reactions with stellar wind particles, can drive this
gas to further radii is uncertain.  However, understanding the dynamics
of the diffuse gas at large radii could be important for interpreting
absorption line studies of transiting systems.

\section{Conclusions}

We have presented two-dimensional, hydrodynamic calculations of the
steady-state structure of thermally driven winds from highly irradiated,
close-in EGPs, including the interaction with a high-velocity wind from
the central star.  Our primary conclusions are that
\begin{itemize}
\item the mass loss rate in an anisotropic wind is reduced by about a factor
of four in a nearly isothermal wind ($\gamma=1.01$) compared to a spherical
wind with the same parameters,
\item the sonic point in an anisotropic wind on the day-side
is located 2-3 times closer to the planet surface 
compared to a spherically symmetric wind with the same parameters,
\item a supersonic polar flow from the day-side to the night-side is
generated just above the planet surface (within $r/R_{p}\approx 2$).  This flow
generates weak shocks due to geometrical compression on the night side.
At large radial distances, the outflow is nearly spherically symmetric.
\item the interaction with a stellar wind strongly compresses the planet wind
and greatly enhances the column density of the gas in the outflow surrounding
the planet.  However, the termination shock in the planetary wind is located
well above the sonic surface, so the overall mass loss rate is affected very 
little.
\end{itemize}

There are a number of limitations to our results.  Probably the most
important is the very simplistic treatment of the thermodynamics in the
wind that we have adopted in our computations.  Rather than attempting to
model the heating, cooling, and thermal conduction processes in the wind
directly, we have simply fixed the temperature at the base of the wind
consistent with previous more detailed models, and computed polytropic
models with different $\gamma$.  However, by comparing our models to
spherically symmetric solutions computed using the same techniques and
parameters, we are able to isolate the effects that multidimensional
dynamics have on the wind.  We conclude these effects can be important,
and therefore it is desirable to include a more realistic treatment of
the radiative transfer and microphysics (e.g. Murray-Clay et al. 2008)
in multidimensional models in the future.

Our conclusions are mostly applicable to winds in which the hydrodynamic
escape parameter $\lambda_{0}=5$, a value which is three times smaller
than that inferred for observed systems such as HD209458b (assuming
the temperature at the base of the wind is $T=10^{4}$~K, and the wind
is mostly neutral).  A slightly higher temperature, or an ionized wind
would produce a value closer to that adopted in most of our models.
For larger values of $\lambda_{0}$, we find the sonic point is located
much farther from the planet, and polar flows produce a spherically
symmetric, extended atmosphere above the planet which serves as the base
of a nearly spherically symmetric wind.  However, the assumptions we have
adopted for the thermodynamics in the wind are likely not applicable in
this case.  Thus, the study if the multidimensional structure of winds
in this parameter regime will require more sopisticated treatments of
the radiative heating and cooling.

Additional limitations to our calculations are that we have not included
the orbital motion of the planet, or the gravitational field of the
central star.  We also have not included the effects of magnetic fields
that are likely to be important in the dynamics of the stellar wind.
Finally, non-thermal plasma processes (such as charge exchange reactions)
might be important in the extremely low density wind material far from the
planet, and since the terminal velocity of the planet winds studied here
is less than the orbital velocity of the planet, these processes might
even dominate the escape of gas from the system.  These limitations
suggest that fully 3D MHD models of anisotropic winds from EGPs are
warranted, and that ideally these models would span the entire orbital
plane of the planet, include possibly important kinetic plasma effects,
and would be based on an improved treatment of the thermodynamics in
the wind.

\acknowledgements
We thank the late Bohdan Paczynski, who was an inspiration for the
work presented in this paper.  We also thank R. Kurosawa for making
figure 8, and A. Burrows, R. Murray-Clay, and R. Rafikov for helpful comments
on an early draft.
D. P. acknowledges support by the National Aeronautics and Space
Administration under Grant/Cooperative Agreement No. NNX08AE57A issued
by the Nevada NASA EPSCoR program.  JS thanks the IAS for support as
a summer visitor.



\begin{figure}
\begin{picture}(340,450)
\put(200,70){\includegraphics{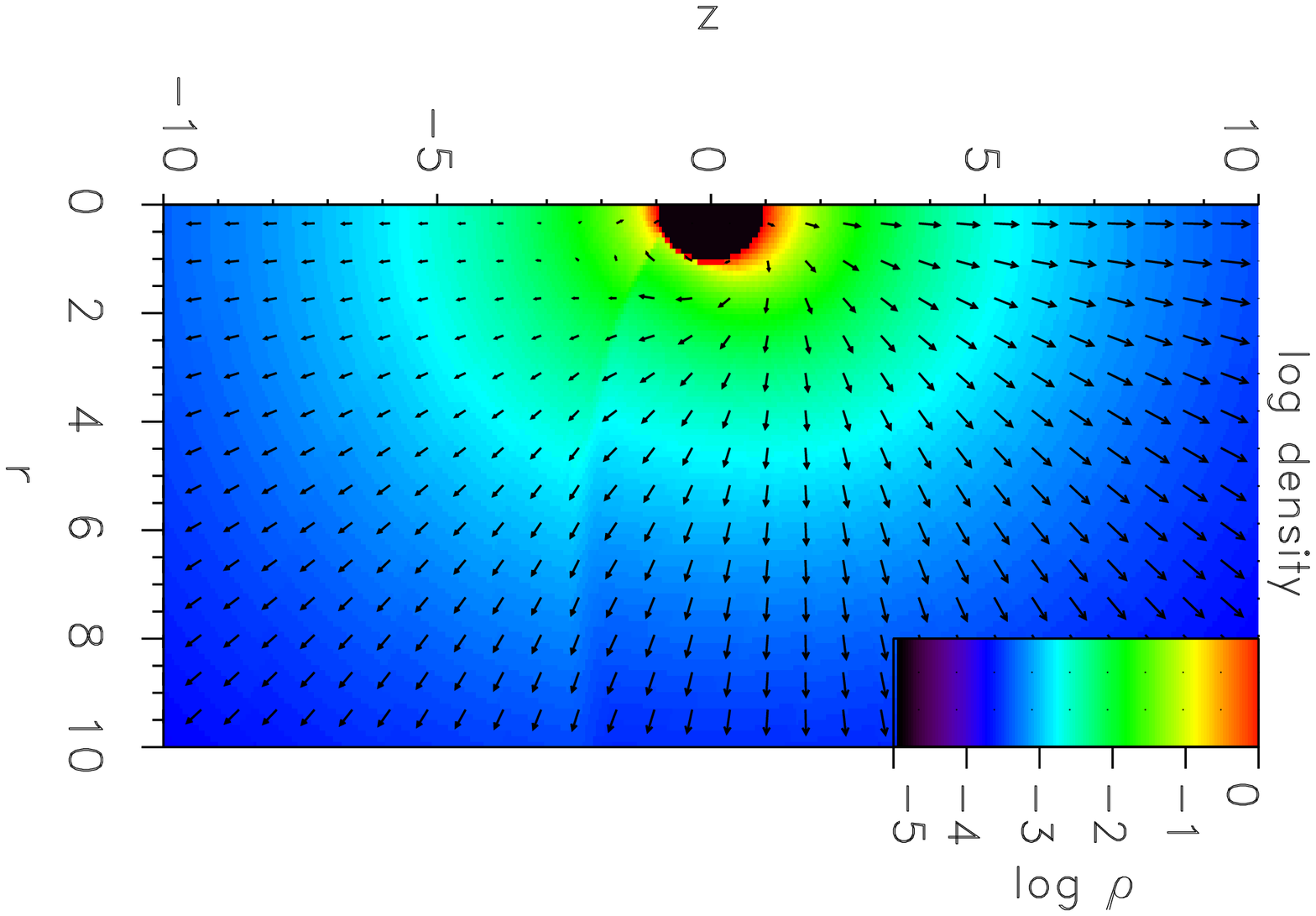}}
\put(400,70){\includegraphics{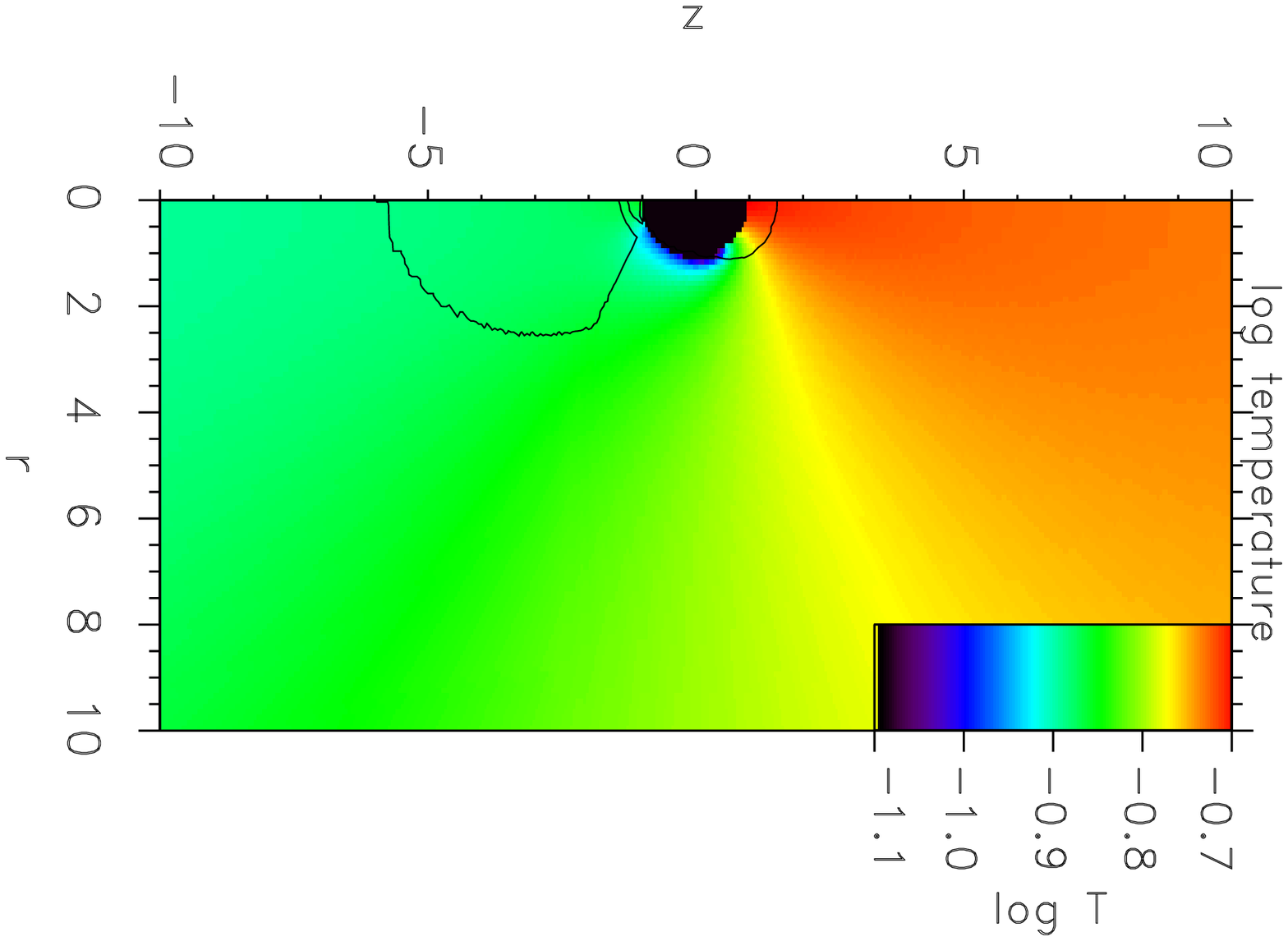}}
\end{picture}
\caption{
{\em Left.} Density (in units of $\rho_{0}$, the value at the base of
the wind) and velocity vectors (in units of the Keplerian velocity at $R_{p}$)
in a nearly isothermal
($\gamma=1.01$) thermally driven wind from a close-in EGP irradiated
on one side only.  {\em Right.}  Temperature, in units of 
$GM_{p}\mu/(k R_{p})$, 
for the same calculation.  The solid line in the temperature plot shows 
the sonic
surfaces.  The irradiating star is located towards the top (corresponding
to $\theta=0$).  } 
\end{figure}

\begin{figure}
\begin{picture}(280,540)
\put(50,160){\includegraphics{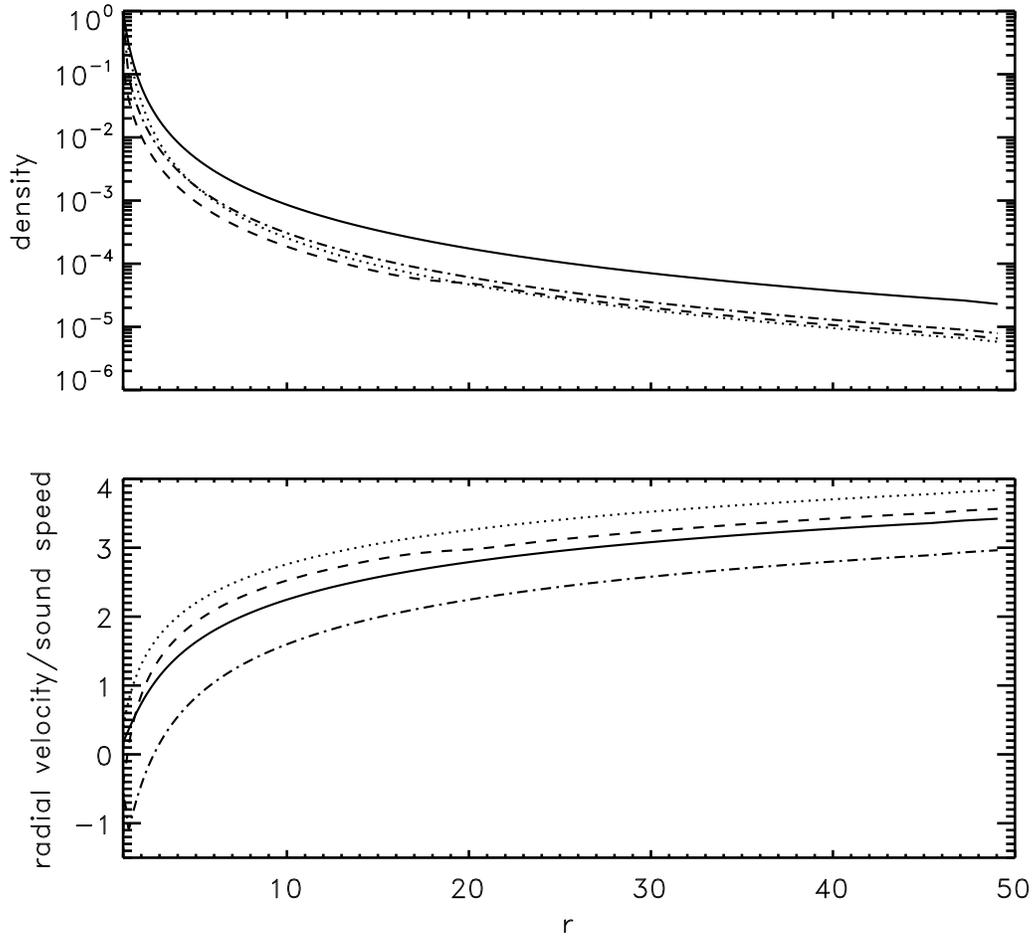}}
\end{picture}
\caption{
Radial profiles of the density and radial velocity (normalized by the
sound speed) at $\theta=0$ (dotted line), $\pi/2$(dashed line), and
$\pi$ (dot-dashed line) for the solution shown in figure 1.  The star
is located at $\theta=0$.  The solid line in each panel shows the profile
for a spherically symmetric wind with $\lambda_{0}=5$.  Radius is measured in units of $R_{p}$.  }
\end{figure}

\begin{figure}
\begin{picture}(280,540)
\put(50,200){\includegraphics{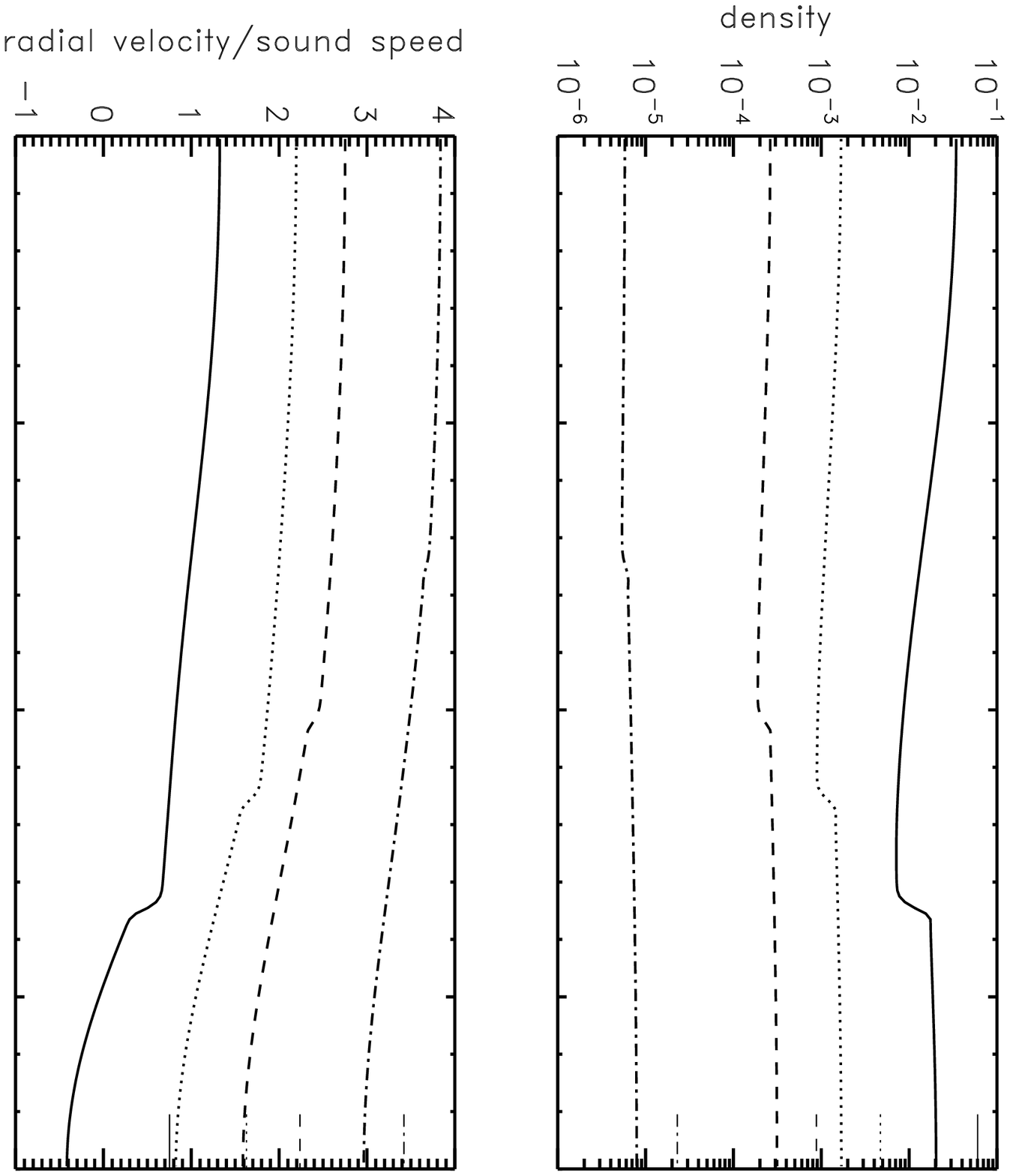}}
\put(50,-150){\includegraphics{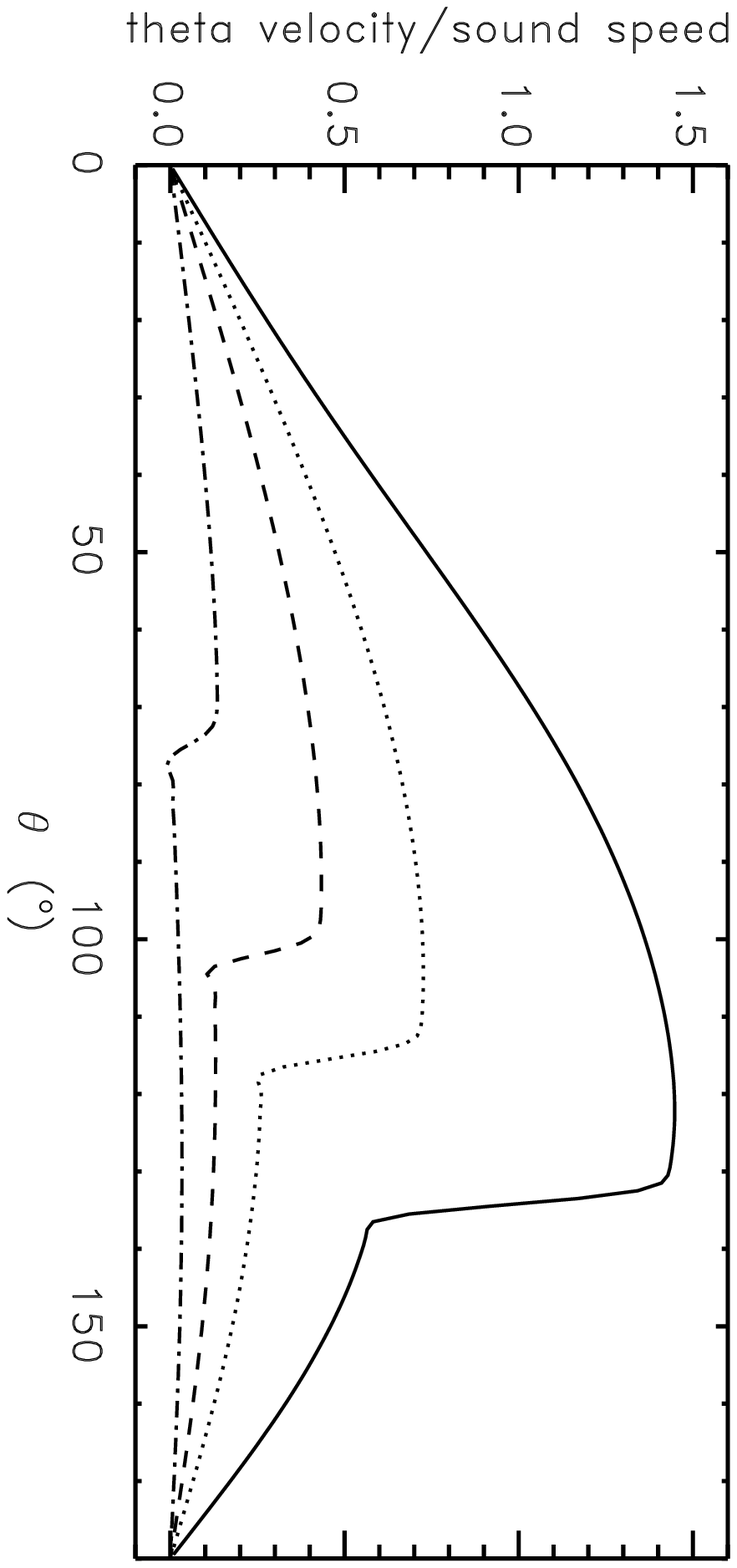}}
\end{picture}
\caption{
Angular profiles of the density, radial and angular velocity (both scaled
by the sound speed) at $r/R_{p}=2$ (solid line), 5 (dotted line), 10
(dashed line), and 50 (dot-dashed line).  The horizontal tick marks at
the right edge of the top two plots show the amplitude of the spherical
wind solution with $\lambda_{0}=5$ at the same radial locations.  }
\end{figure}

\begin{figure}
\begin{picture}(340,450)
\put(200,70){\includegraphics{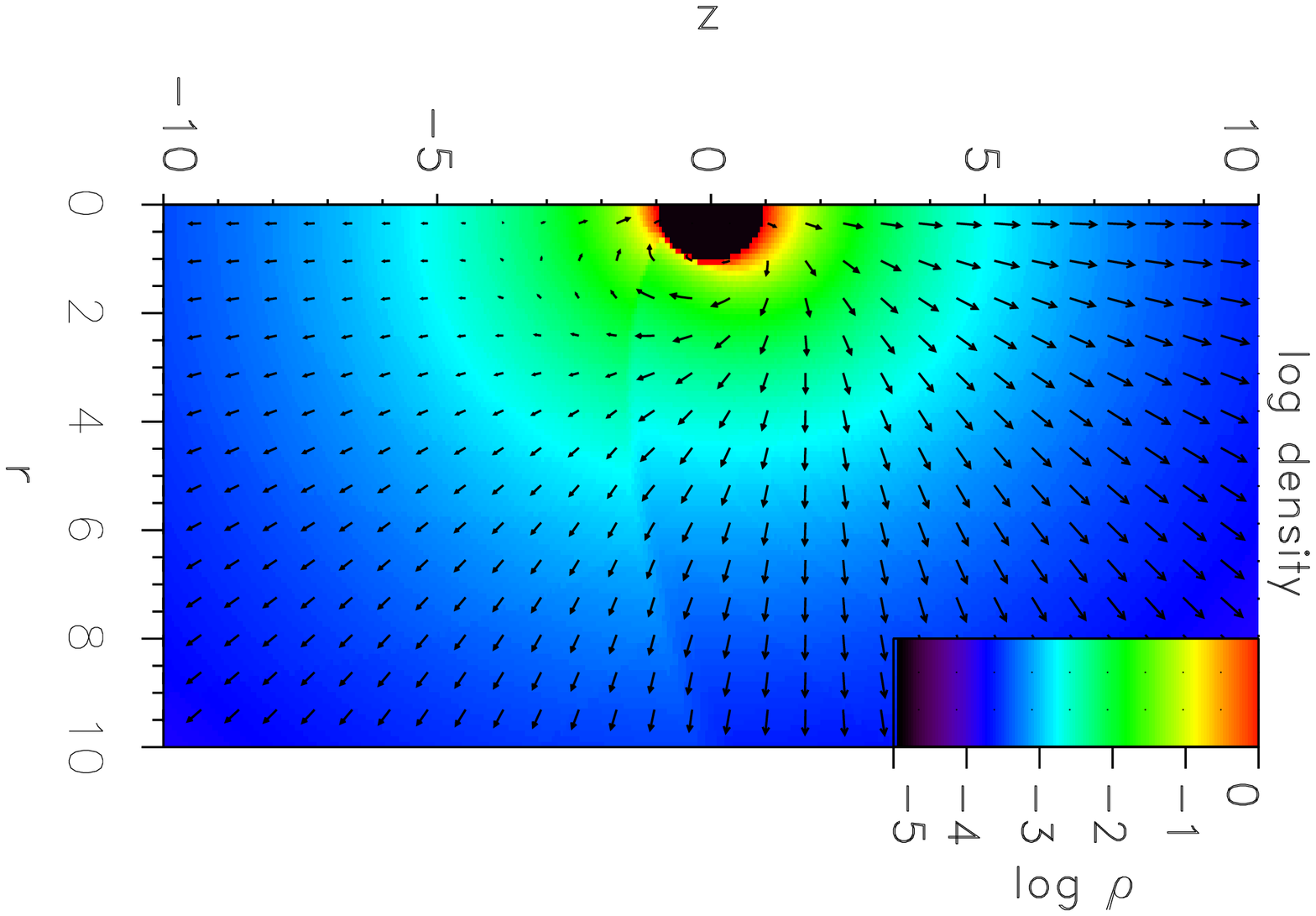}}

\put(400,70){\includegraphics{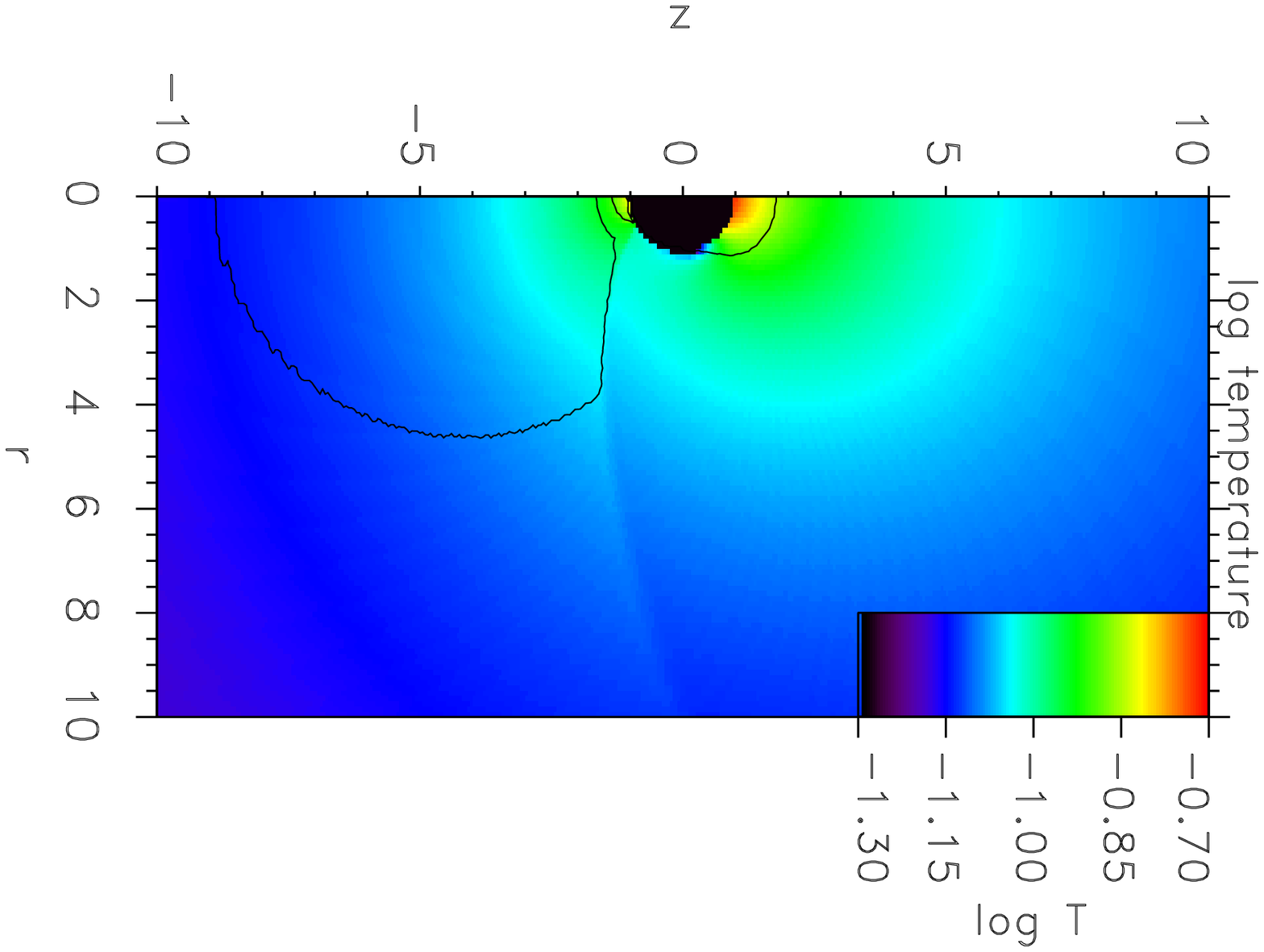}}
\end{picture}
\caption{
{\em Left.} Density (in units of $\rho_{0}$, the value at the base of
the wind) and velocity vectors (scaled to the Keplerian velocity at $R_{0}$) 
for a anisotropic
EGP wind with $\gamma=1.1$.
{\em Right.}  Temperature, in units of 
$GM_{p}\mu/(k R_{p})$,
for the
same calculation.  The solid line in the temperature plot shows the sonic
surfaces.  The irradiating star is located towards the top.  } 
\end{figure}

\begin{figure}
\begin{picture}(280,540)
\put(50,160){\includegraphics{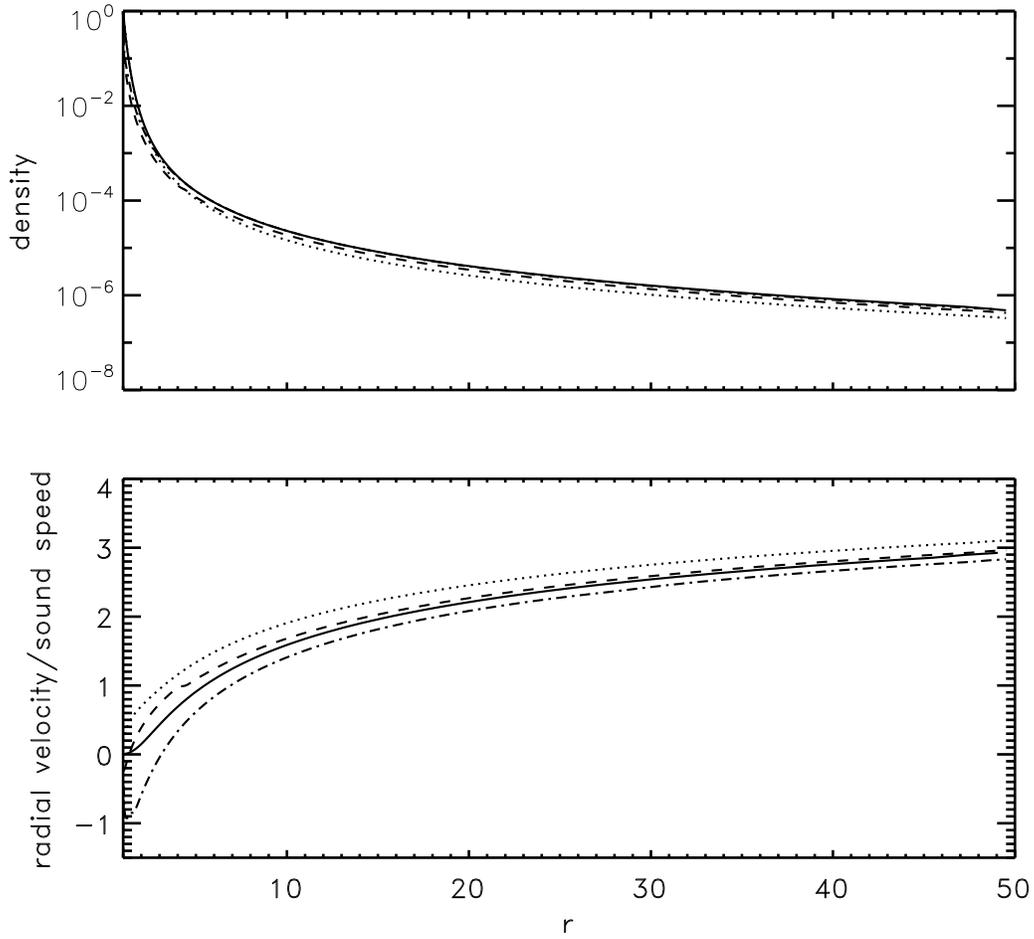}}
\end{picture}
\caption{
Radial profiles of the density and radial velocity (normalized by the
sound speed) at $\theta=0$ (dotted line), $\pi/2$(dashed line), and
$\pi$ (dot-dashed line) for a model with $\lambda_{0}=10$.  The star
is located at $\theta=0$.  The solid line in each panel shows the profile
for a spherically symmetric wind with the same $\lambda_{0}$.  Radius is measured in units of $R_{p}$.  }
\end{figure}

\begin{figure}
\begin{picture}(340,450)
\put(200,70){\includegraphics{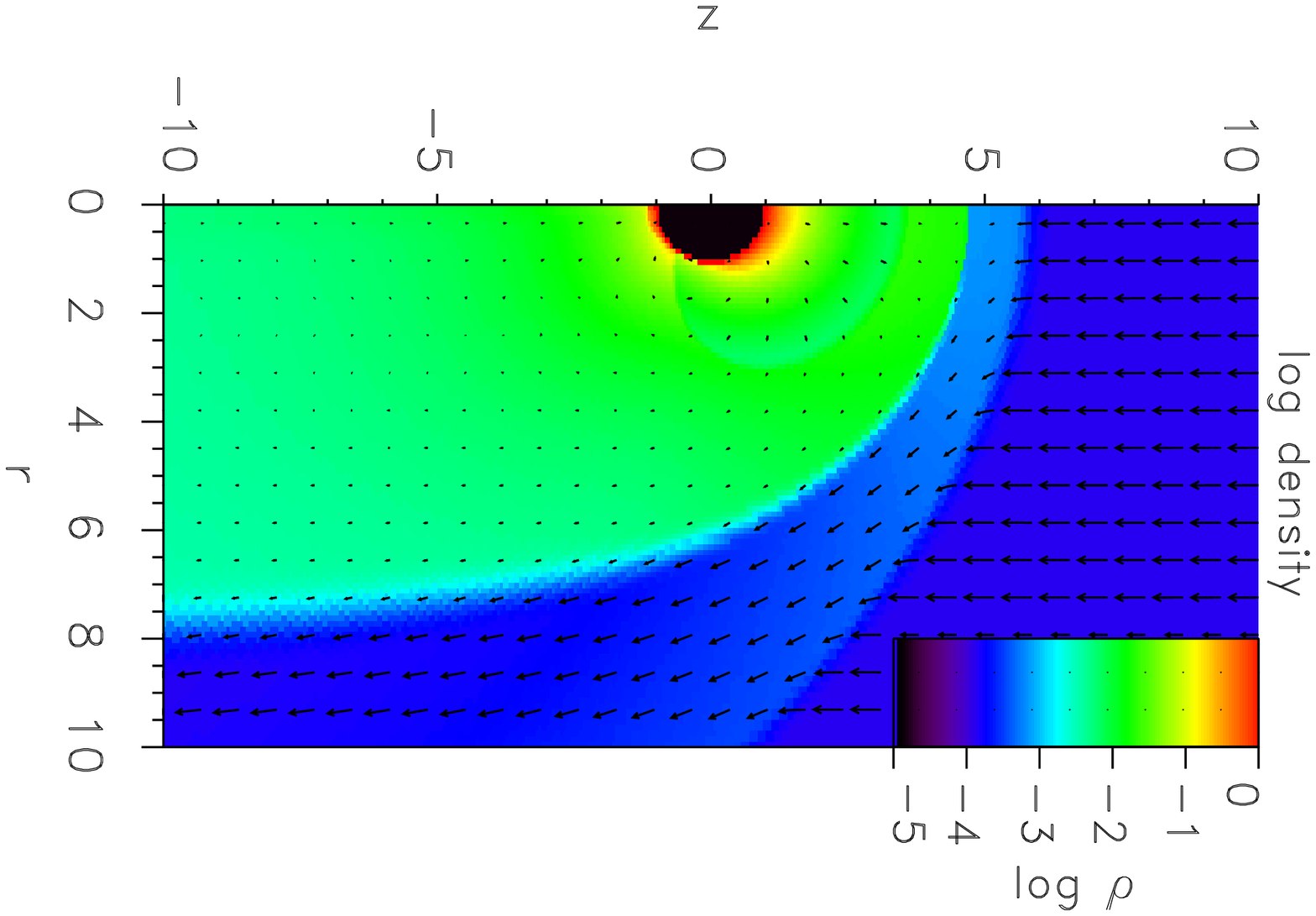}}

\put(400,70){\includegraphics{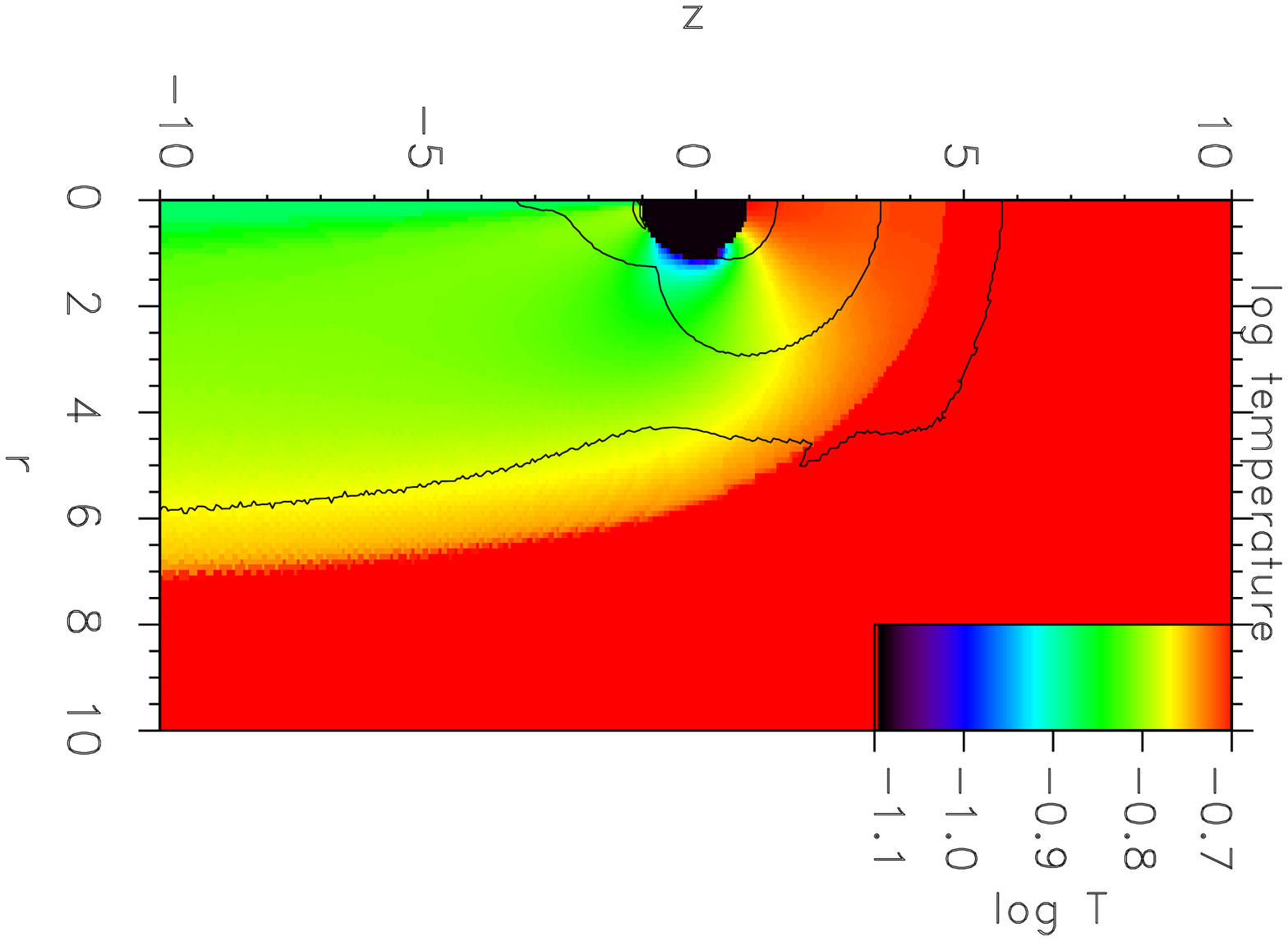}}
\end{picture}
\caption{
({\em Left.}) Density (in units of $\rho_{0}$, the value at the base of
the wind) and velocity vectors (scaled to the Keplerian velocity at $P_{0}$) 
for a anisotropic
EGP wind with $\gamma=1.01$ interacting with a uniform stellar wind.
({\em Right.}) Temperature in units of 
$GM_{p}\mu/(k R_{p})$,
for the
same calculation.  The solid line in the temperature plot shows the sonic
surfaces.  The star is located towards the top.
}
\end{figure}

\begin{figure}
\begin{picture}(280,540)
\put(50,160){\includegraphics{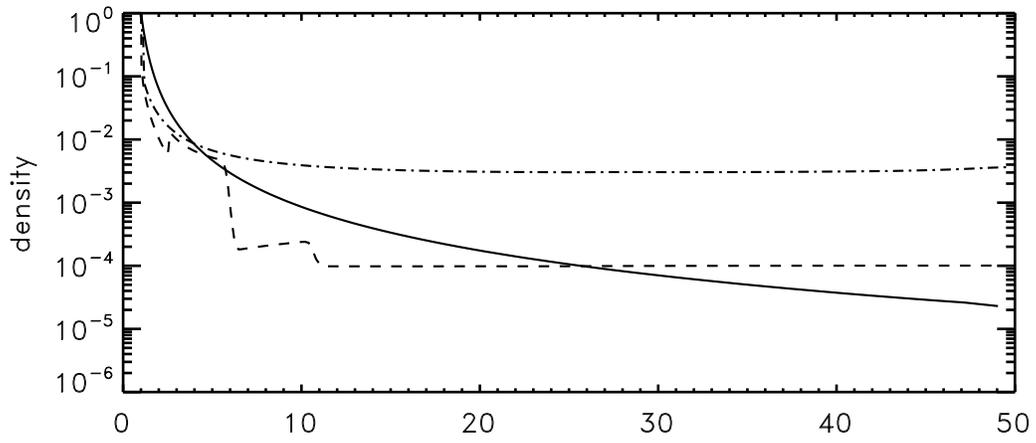}}

\end{picture}
\caption{Radial profiles of the density at $\theta=\pi/2$
(dashed line), and $\pi$ (dot-dashed line) for a planetary wind interacting
with a stellar wind.  The solid line
shows the profile in a spherically symmetric wind.  Radius is measured in units of $R_{p}$.
}
\end{figure}

\begin{figure}
\begin{picture}(280,540)
\put(50,160){\includegraphics{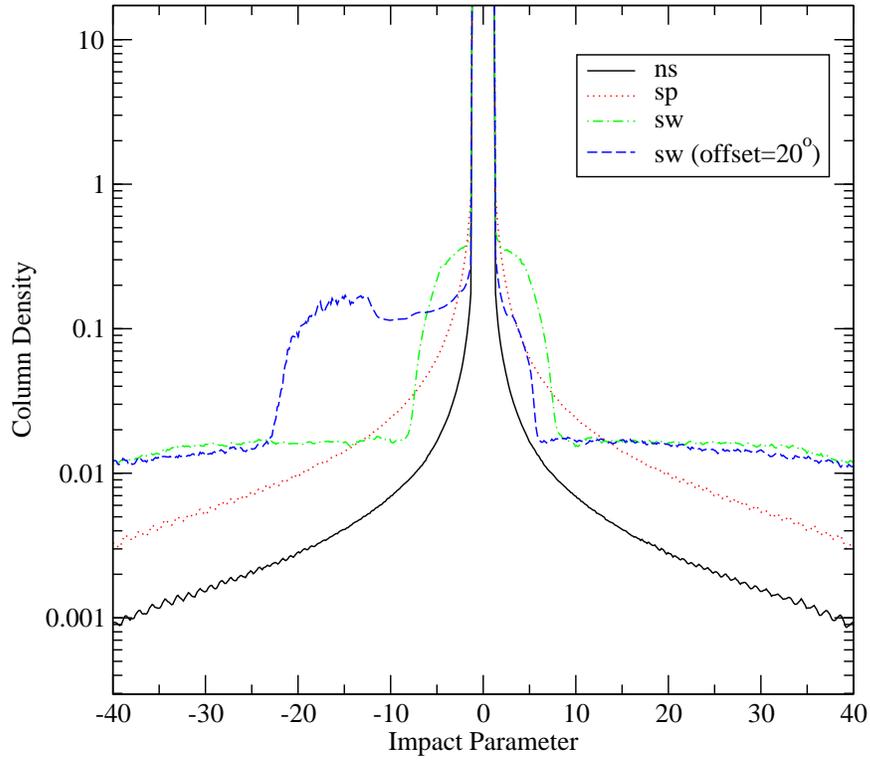}}

\end{picture}
\caption{Column density (in arbitrary units) as a function of impact parameter (in units
of $R_{p}$) for four different models of thermally driven hydrodynamic winds from
a close-in EGP.  The dotted line (labeled `sp') corresponds to a
spherically symmetric wind.  The solid line (labeled `ns') corresponds to an
anisotropic planetary wind that does not interact with a stellar wind.
The dot-dashed line (labeled `sw') corresponds to
an anisotropic wind interacting with a stellar wind, with the observer
located at an angle of $\theta=\pi$.  The dashed line is the same as `sw', but
with the observer located at an angle offset by $20^{\circ}$ from $\theta=\pi$.
}
\end{figure}


\begin{references}

\reference{} Ballester, G.E., Sing, D.K., \& Herbert, F., 2007.  Nature, 445, 511.
\reference{} Ben-Jaffel, L. 2007.  ApJ, 671, L61.
\reference{} Borovikov, S.N., Pogorelov, N.V., Zank, G.P., \& Kryukov, I.A. 2008.  ApJ, 682, 1404.
\reference{} Dobbs-Dixon, I., \& Lin, D.N.C., 2008.  ApJ, 673, 513.
\reference{} Ehrenreich, D., 2008.  available at arXiv:0807.1885v1.
\reference{} Ehrenreich, D., et al. 2008.  A\&A 483, 933.
\reference{} Erkaev, N.V., et al. 2007.  A\&A 472, 329.
\reference{} Garc\'{i}a Mu\~{n}oz, A. 2007.  P\&SS 55, 1426.
\reference{} Hubbard, W.B., Hattori, M.F., Burrows, A., Hubeny, I., \& Sudarsky, D., 2007.  Icarus, 187, 358.
\reference{} Hunten, D.M. 1982.  P\&SS, 30, 773
\reference{} Keppens, R., \& Goedbloed, J.P., 1999.  A\&A 343, 251.
\reference{} Lammer, H., Selsis, F., Ribas, I., Guinan, E.F., Bauer, S.J., \& Weiss, W.W., 2003.  ApJ 598, L121.
\reference{} Lecavelier des Etangs, A., 2007.  A\&A 461, 1185.
\reference{} Lecavelier des Etangs, A., Vidal-Madjar, A., McConnell, J.C.,
\& H\'{e}brard, G., 2004.  A\&A 418, L1.
\reference{} Murray-Clay, R., Chiang, E., \& Murray, N., ApJ in press (arXiv:0811.0006).
\reference{} Moutou, C., et al. 2001.  A\&A 371, 260.
\reference{} Preusse, S., Kopp, A., B\"{u}chner, J., \& Motschmann, U., 2005.  A\&A 434, 1191.
\reference{} Proga, D., Stone, J.M., \& Drew, J.E., 1998.  MNRAS 295, 595.
\reference{} Proga, D., Stone, J.M., \& Kalman, T.R., 2000.  ApJ 543, 686.
\reference{} Schneider, J., 2008.  The Extrasolar Planets Encyclopedia, {\tt http://exoplanets.eu}
\reference{} Schneiter, E.M., Vel\'{a}zquez, P.F., Esquivel, A., \& Raga, A.C. 2007.  ApJ 671, L57.
\reference{} Showman, A.P., Cooper, C.S., Fortney, J.J., \& Marley, M.S., ApJ 682, 559.
\reference{} Stone, J.M. \& Norman, M.L. 1992.  ApJS, 80, 753.
\reference{} Tian, F., Toon, O.B., Pavlov, A.A., \& De Sterck, H., 2005.  ApJ 621, 1049.
\reference{} Vidal-Madjar, A., et al. 2003.  Nature, 422, 143.
\reference{} Vidal-Madjar, A., et al. 2008.  ApJ, 676, L57.
\reference{} Watson, A.J., Donahue, T.M., \& Walker, J.C.G., 1981.  Icarus, 48, 150.
\reference{} Yelle, R.V. 2004.  Icarus, 170, 167.  Also corrigendum, Icarus, 183, 508.

\end{references}
\end{document}